\documentclass[12pt,preprint]{aastex}
\usepackage{natbib}

\newcommand{\kms}{$\rm km\;s^{-1}$}  
\newcommand{\vsigma}{$V_c$--$\,\sigma_c$}
\newcommand{\sigmac}{$\sigma_c$}
\newcommand{\Vc}{$V_c$}
\newcommand{\Vvir}{$V_{\it vir}$}
\newcommand{\hi}{H~{\small I}} 
 
\shorttitle{\vsigma\ relation in HSB and LSB galaxies}
\shortauthors{Pizzella et al.}

\begin{document}


\title{On the relation between circular velocity and central velocity
dispersion in high and low surface brightness galaxies\thanks{Based on
observations made with European Southern Observatory Telescopes at the
Paranal Observatory under programmes 67.B-0283, 69.B-0573 and
70.B-0171.
\/}}


\author{A. Pizzella\altaffilmark{1}, 
E.M. Corsini\altaffilmark{1}, 
E. Dalla Bont\`a\altaffilmark{1},
M. Sarzi\altaffilmark{2},
L. Coccato\altaffilmark{1}}

\and 

\author{F.~Bertola\altaffilmark{1}}

\altaffiltext{1}{Dipartimento di Astronomia, Universit\`a di Padova, 
vicolo dell'Osservatorio 2, I-35122 Padova, Italy}

\altaffiltext{2}{Physics Department, University of Oxford, Keble Road, 
Oxford, OX1 3RH, UK}


\begin{abstract}
In order to investigate the correlation between
the circular velocity \Vc\ and the central velocity dispersion of the
spheroidal component \sigmac , we analyzed these quantities for a
sample of 40 high surface brightness disc galaxies (hereafter HSB),
8 giant low surface brightness spiral galaxies (hereafter LSB),
and 24 elliptical galaxies characterized by flat rotation curves.
  Galaxies have been selected to have a velocity gradient $\leq2$ 
$\rm km\;s^{-1}\;kpc^{-1}$ for $R  \geq 0.35  R_{25}$. 
We used these data to better define the previous \vsigma\ correlation for
spiral galaxies (which turned out to be HSB) and elliptical galaxies,
especially at the lower end of the \sigmac\ values.
We find that the \vsigma\ relation is described by a linear law out to
velocity dispersions as low as \sigmac$\approx50$
\kms, while in previous works a power law was adopted
for galaxies with \sigmac$>80$ \kms.

Elliptical galaxies with \Vc\ based on dynamical models or directly
derived from the \hi\ rotation curves follow the same relation as the
HSB galaxies in the \vsigma\ plane. On the contrary, the LSB galaxies
follow a different relation, since most of them show either higher
\Vc\ (or lower \sigmac ) with respect to the HSB galaxies.
This argues against the relevance of baryon collapse in the radial
density profile of the dark matter haloes of LSB galaxies.
Moreover, if the \vsigma\ relation is equivalent to one between the
mass of the dark matter halo and that of the supermassive black hole,
these results suggest that the LSB galaxies host a supermassive black
hole with a smaller mass compared to HSB galaxies of equal dark matter
halo.
  On the other hand, if the fundamental
correlation of SMBH mass is with the halo circular velocity, then LSBs
should have larger black hole masses for given bulge dispersion.  

Elliptical galaxies with \Vc\ derived from \hi\ data and LSB galaxies
were not considered in previous studies.
\end{abstract}




\section{Introduction}
\label{sec:introduction}

A possible relation between the central velocity dispersion of
the spheroidal component (\sigmac ) and the galaxy circular velocity
measured in the flat region of the rotation curve (\Vc ) was suggested
by \citet{Whit1979} and \citet{Whit1981}. By measuring \hi\ line
widths they found that \Vc$/$\sigmac\ $\sim 1.7$ for a sample of S0
and spiral galaxies. Recently, \citet{Ferr2002} proceeded further
extending the \vsigma\ relation to elliptical galaxies.
She interpreted the \vsigma\ relation as suggestive of
a correlation between two different galactic components, since \sigmac\
and \Vc\ probe the potential of the spheroidal component and of the dark
matter (hereafter DM) halo, respectively.
In particular, it results that for a given DM halo the central
velocity dispersion of the spheroidal component is independent of the
morphological type.
The validity of this relation has been confirmed by \citet{Baes2003},
who enlarged the sample of spiral galaxies.

For elliptical galaxies, \Vc\ is generally inferred by means of
dynamical modelling of the stellar kinematics.  This is the case of
the giant round and almost non-rotating ellipticals studied by
\citet{Kron2000}. 
  These galaxies form a unique dynamical family which
scales with luminosity and effective radius. 
As a consequence
the maximum circular velocity is correlated to the central
velocity dispersion of the galaxy. 
\citep{Gerh2001}. Whether the same is true for more flattened and
fainter ellipticals is still to be investigated.
On the contrary, both shape and amplitude of the rotation curve of a
spiral galaxy depend on the galaxy luminosity and morphological type
\citep[e.g.,][]{Burs1985, pers1996}. For this reason for spiral
galaxies the \vsigma\ relation is not expected a priori.

It is interesting to investigate whether the \vsigma\ relation holds
also for less dense objects characterized by a less steep potential
well. This is the case of low surface brightness galaxies (hereafter
LSB), which are disc galaxies with a central face-on surface
brightness $\mu_B\geq22.6$ mag arcsec$^{-2}$
\citep[e.g.,][]{Scho1992,Impe1996}.
  Previous work concentrated on HSB and, to infer the \Vc\ for 
elliptical galaxies, they relied on stellar dynamical models.
In this work we investigated the behavior of elliptical 
galaxies with HI-based \Vc\ and of LSB galaxies in the
\vsigma\ relation. 

This paper is organized as follows. An overview of the properties of
the sample galaxies as well as the analysis of the kinematic data
available in literature to derive their \Vc\ and \sigmac\ are
presented in Sect. \ref{sec:sample}. The results and discussion
concerning the \vsigma\ relation are given in Sect. \ref{sec:results}.
 
\section{Sample selection}
\label{sec:sample}

In the past years we started a scientific program aimed at deriving
the detailed kinematics of ionized gas and stars in HSB and LSB
galaxies in order to study their mass distribution and structural
properties.
We measured the velocity curves and velocity dispersion profiles along
the major axis for both the ionized-gas and stellar components for a
preliminary sample of $50$ HSB galaxies [$10$ S0--S0/a galaxies in
\citet[]{Cors2003}; $7$ Sa galaxies in \citet[]{Bert1996} and
\citet[]{Cors1999}; $16$ S0--Sc galaxies in \citet[]{vega2001}; $17$
Sb--Scd galaxies in \citet[]{Pizz2004..17spirals}] and 11 LSB galaxies
\citep[]{Pizz2004..LSB, Pizz2004IAUS}

The HSB sample consists of disc galaxies with Hubble type ranging from
S0 to Scd, an inclination $i \geq 30^\circ $ and a distance $D < 
80$ Mpc.
The LSB sample consists of disc galaxies with Hubble type ranging
from Sa to Irr, an intermediate inclination ($30^\circ \la i <
70^\circ$), and a distance $D < 65$ Mpc (except for ESO
534-G20). Three LSB galaxies, namely ESO 206-G14, ESO 488-G49,
and LSBC F563-V02, have been selected from the sample observed by
\citet{Debl1997}. The remaining eight objects are LSB galaxies with
bulge. They have been selected in Lauberts \& Valentijn (1989,
hereafter ESO-LV) to have a LSB disc component following the criteria
described by \citet{Beij1999}. Due to the bulge light contribution the
total central face-on surface brightness of the galaxy could be
$\mu_B\leq22.6$ mag arcsec$^{-2}$.  However, all these objects do
have a LSB disc. As far as their total luminosity concerns, LSBC
F563-V02 and ESO 488-G49 are two dwarf LSB galaxies but the other nine
objects are representative of giant LSB galaxies
\citep[e.g.,][]{Mcga2001}.

For all the HSB and LSB galaxies we obtained the ionized-gas rotation
curve by folding the observed line-of-sight velocities around the
galaxy center and systemic velocity after averaging the contiguous
data points and applying a correction for galaxy inclination.  We
rejected $35$ HSB galaxies because they had asymmetric rotation curves
or rotation curves which were not characterized by an outer flat
portion. 
 \citet{Ferr2002} and
\citet{Baes2003} 
considered galaxies with the rotation curve extending farther out than
$R_{25}$. This criterion is not appropriate when the sample galaxies
spans a wide range in photometrical properties. For LSB galaxies,
which have a lower central surface brightness than HSB galaxies,
$R_{25}$ corresponds to a relatively small radius where the rotation
curve may be still rising. For this reason we adopt a criterion that
select rotation curves on the basis of their flatness rather then on
their extension.  

\begin{figure}
\epsscale{.80}
\plotone{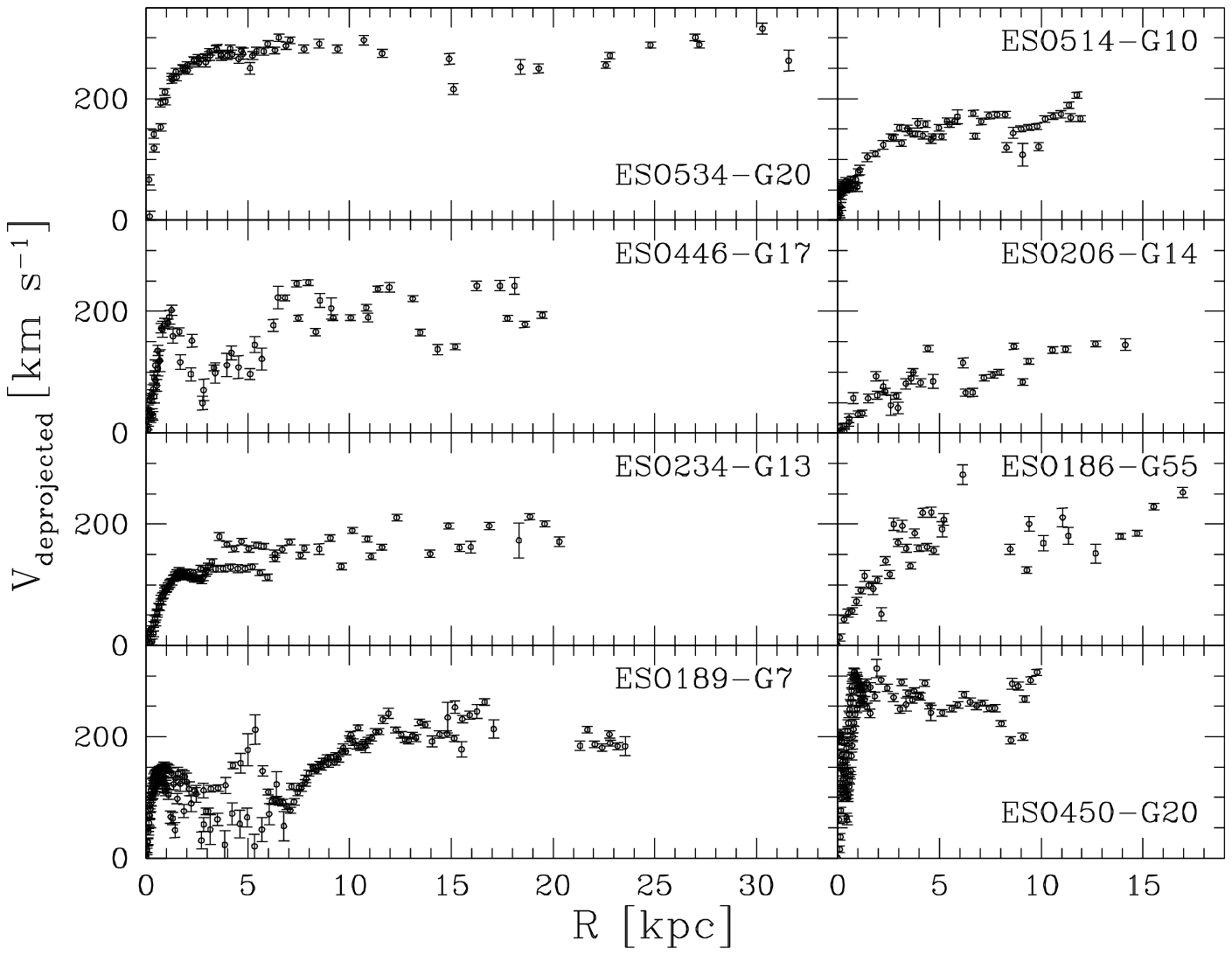}
\caption{Deprojected rotation curves of the eight LSB galaxies of the final sample.}
\label{fig:LSBvelcur}
\end{figure}

The  flatness of each rotation curve  has been checked by fitting it
with a linear  low $V(R)=AR+B$  for $R  \geq 0.35  R_{25}$. The radial
range has been chosen in order  to avoid the bulge-dominated region of
the rotation curve  (e.g., IC 724 and  NGC 2815). The rotation  curves
with $|A|\geq2$  $\rm km\;s^{-1}\;kpc^{-1}$ within 3$\sigma$ have been
considered not to be flat.  In this way $15$ HSB  galaxies and $8$ LSB
galaxies resulted  to have a flat rotation  curve.  Since the velocity
curves of the  LSB galaxies were  not presented in previous papers, we
show their folded rotation curves in Fig. \ref{fig:LSBvelcur}.
We derived \Vc\ by averaging the outermost values of the flat portion
of the rotation curve.
 
We are therefore confident that we are giving a reliable estimate of
the asymptotic value of the circular velocity which traces the mass of
the DM halo \citep[see][ for a discussion]{Ferr2002}.
We derived \sigmac\ from the stellar kinematics by extrapolating the
velocity dispersion radial profile to $r=0''$.  This has been done by
fitting the $8$ innermost data points with an empirical function (either
an exponential, or a Gaussian or a constant). We did not apply any
aperture correction to \sigmac\ as discussed by \citet{Baes2003},
and \citet{Pizz2004..17spirals}.

In order to complete our sample of disc galaxies we included all the
spiral galaxies previously studied by \citet{Ferr2002} and
\citet{Baes2003}, but which are in addition
characterized by a flat rotation curve. 
  We therefore applied to this latter galaxy sample 
the same flatness criterion applied to our sample .
  
In summary, we have $23$ galaxies ($15$ HSB and $8$ LSB galaxies) from
our preliminary sample, $16$ spiral galaxies (out of 38) from
\citet{Ferr2002}, and $9$ spiral galaxies (out of 12) from
\citet{Baes2003}. It should be noted that the final sample of HSB
galaxies includes $11$ early-type objects with Hubble type ranging
from S0 to Sab. On the contrary, the sample by
\citet{Baes2003} and \citet{Ferr2002} was constituted only by
late-type spirals with Hubble type Sb or later (except for the Sa NGC
2844).

Finally, we considered a sample of 24 elliptical galaxies with a flat
rotation curve and for which both \Vc\ and \sigmac\ are available from the
literature. They include 19 objects studied by \citet{Kron2000} who
derived \Vc\ by dynamical modeling and 5 objects for which \Vc\ is
directly derived from the flat portion of their \hi\ rotation curves.
The addition of these last 5 ellipticals is important as it allows
to test against model-dependent biases in the $V_c-\sigma_c$
relation.

The \Vc\ of NGC 4278 has been estimated from both the \hi\ rotation curve
\citep[][ at a distance from the center of 3.3 $R_{25}$]{Lees1994} 
and dynamical models \citep[][ at a distance from the center of 0.1
$R_{25}$]{Kron2000}. The values are in agreement within 2$\sigma$
error bars. For the further analysis we adopted the \hi\ \Vc\ which
has been obtained at a larger distance from the center.

The values \sigmac\ of all the elliptical galaxies have been corrected
to the equivalent of an aperture of radius $r_e/8$ following the
prescriptions of \citet{Jorg1995}. The effective radius $r_e$ is taken
from de Vaucouleurs et al. (1991, hereafter RC3).

The basic properties of the complete sample of $40$ HSB
disc galaxies, $8$ LSB spiral galaxies and $24$ elliptical galaxies are
listed in Table \ref{tab:sample} as well as their values of \Vc\ and
\sigmac .

\section{Results and discussion}
\label{sec:results}

\begin{figure}
\epsscale{.80}
\plotone{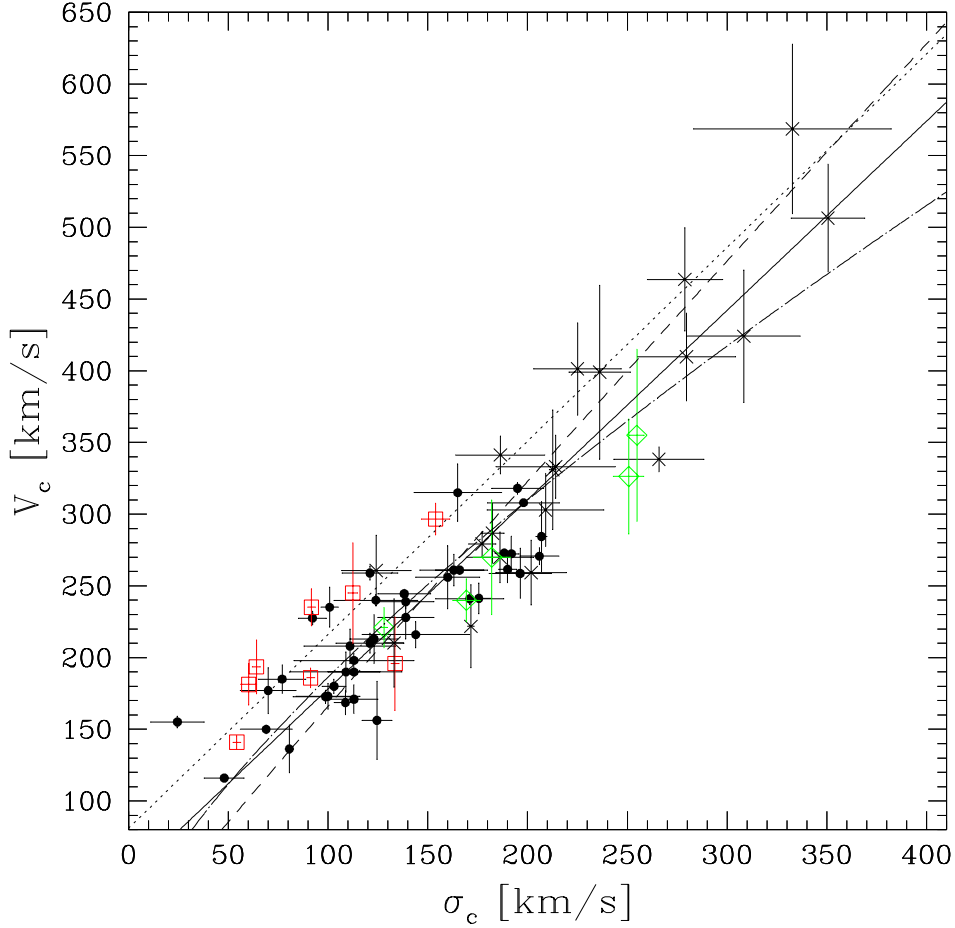}
\caption{The correlation between the circular velocity \Vc\ 
  and the central velocity dispersion of the spheroidal component
  \sigmac\ for elliptical and disc galaxies. The data points
  corresponding to HSB galaxies ({\it filled circles\/}), LSB galaxies
  ({\it squares\/}), elliptical galaxies with \Vc\ obtained from \hi\
  data ({\it diamonds\/}), and elliptical galaxies with \Vc\
  obtained from dynamical models ({\it crosses\/}) are shown. The {\it
  continuous\/} and {\it dash-dotted line\/} represent the linear
  (Eq. \ref{eq:HSBfit_linear}) and power-law fit
  (Eq. \ref{eq:HSBfit_power}) to HSB and elliptical galaxies. The
  {\it dotted line\/} represents the linear-law fit
  (Eq. \ref{eq:LSBfit_linear}) to LSB galaxies. For a comparison,
  the {\it dashed line\/} corresponds to the power-law fit to spiral
  galaxies with \sigmac$>80$ \kms\ by \citet{Baes2003}.}
\label{fig:sample}  
\end{figure}

The \Vc\ and \sigmac\ data points of the final sample of galaxies are
plotted in Fig. \ref{fig:sample}. We applied a linear regression
analysis to the data by adopting the method by \citet[]{Akri1996} for
bivariate correlated errors and intrinsic scatter (hereafter BCES)
both in the $\log V_c$--$\,\log \sigma_c$ and \vsigma\ plane. 
We did not include LSB galaxies in the analysis
because they appear to follow a different \vsigma\ relation as we will
discuss later.

Following Ferrarese (2002) and Baes et al. (2003) we fit the function
$\log V_c = \alpha\log\sigma_c + \beta$ to the data in $\log
V_c$--$\,\log \sigma_c$ plane. We find
\begin{equation}
\log V_c  = (0.74\pm 0.07)\, \log \sigma_c + (0.80 \pm 0.15)  
\label{eq:HSBfit_power}
\end{equation}
with \Vc\ and \sigmac\ expressed in \kms. The resulting power law is
plotted in Fig. \ref{fig:sample}. To perform a comparison with previous
results we defined the reduced $\chi^2$ as in \citet{NUMREC}

\begin{equation}
\chi^2_\nu = \frac{1}{N-2}\sum_{i=1}^N \frac{(\log V_{c,i} - 
\log V^{\it fit}_{c,i})^2}{\Delta \log V_{c,i}^2+\alpha^2\Delta \log \sigma_{c,i}^2}
\end{equation}
where $\Delta \log \sigma_{c,i}$ and $\Delta \log V_{c,i}$ are the
errors of the $i-$th data point, $\log V_{c,i}$ and $\log V^{\it
fit}_{c,i}$ are the observed and fitted velocity of the $i-$th data
point, $\alpha=0.74$ is the linear coefficient of the regression, and
$N=64$ is the number of data points.  
We find $\chi^2_\nu =2.5$.

The fitting power law has $\alpha\approx1$ in agreement with
\citet{Ferr2002} and \citet{Baes2003}. The power-law 
fit by \citet{Baes2003} is plotted in Fig. \ref{fig:sample} for a
comparison. However, \citet{Ferr2002} and \citet{Baes2003} included in
their fits only galaxies with \sigmac $>70$ \kms\ and \sigmac $>80$
\kms , respectively. In fact, they considered the few objects with
\sigmac $\leq70$ \kms\ as outliers.  On the contrary, we found that
points characterized by \sigmac $\la70$ \kms\ appear to be well
represented by the fitting law as well as the ones characterized by
higher values of \sigmac .

Since it results $\alpha\approx1$, we decided to fit the function $V_c
= a \sigma_c + b$ to the data in the \vsigma\ plane.  We find
\begin{equation}
V_c = (1.32\pm 0.09)\,\sigma_c + (46\pm14)
\label{eq:HSBfit_linear}
\end{equation}
with \Vc\ and $\sigma_c$ expressed in \kms. The resulting straight
line is plotted in Fig. \ref{fig:sample}. We find $\chi^2_\nu =2.7$ by 
  defining the reduced $\chi^2$ as
\begin{equation}
\chi^2_\nu = \frac{1}{N-2}\sum_{i=1}^N \frac{(V_{c,i}-V^{\it fit}_{c,i})^2}{\Delta V_{c,i}^2+a^2\Delta\sigma_{c,i}^2}
\end{equation}
where $\Delta \sigma_{c,i}$ and $\Delta V_{c,i}$ are the errors of the
$i-$th data point, $V_{c,i}$ and $V^{\it fit}_{c,i}$ are the observed
and fitted velocity for $i-$th data point, $\alpha=1.35$ is the linear
coefficient of the regression, and $N=64$ is the number of data
points.  .

To summarize, in previous works a power law was adopted to describe the correlation
between \Vc\ and \sigmac\ for galaxies with
\sigmac$>80$ \kms . We find that data are also consistent with a linear
law out to velocity dispersions as low as \sigmac$\approx50$
\kms . We considered the straight line given in Eq. \ref{eq:HSBfit_linear} as
reference fit.

\begin{figure} 
\epsscale{.80}
\plotone{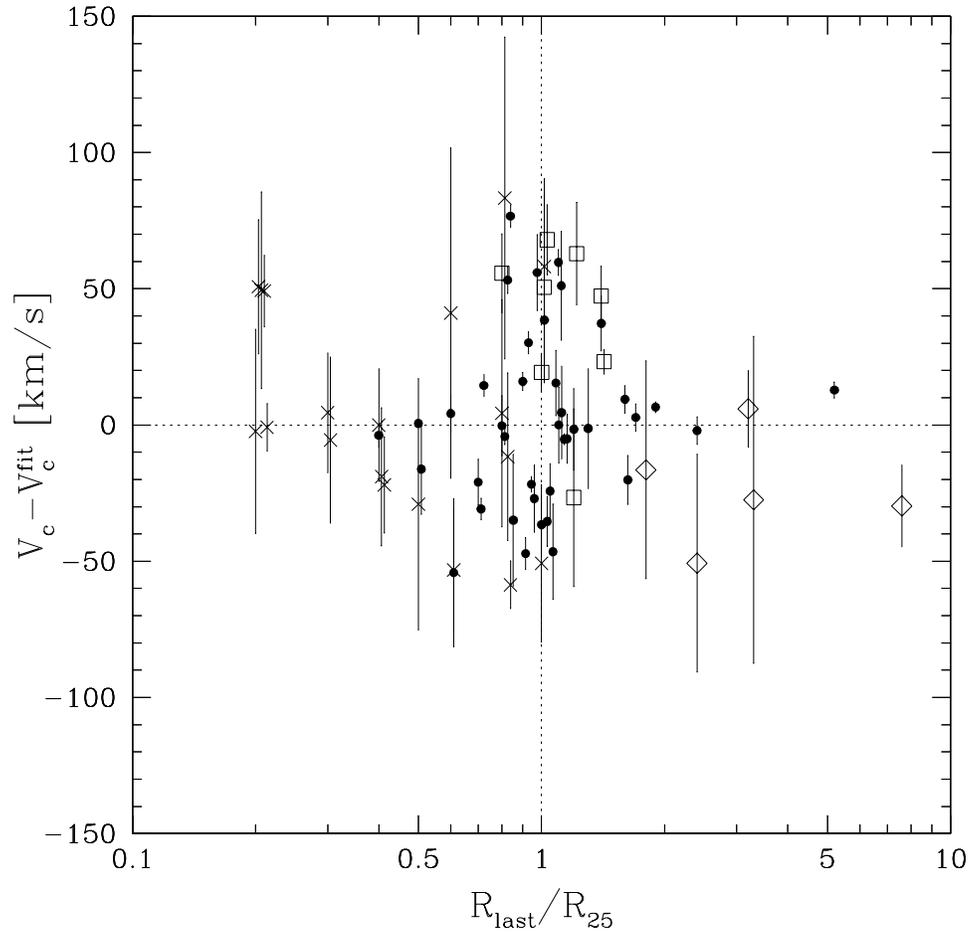}
\caption{Residuals from the linear-law fit to
  HSB and elliptical galaxies (Eq. \ref{eq:HSBfit_linear}) plotted
  as function of $R_{\it last}/R_{25}$.  The data points corresponding to
  HSB galaxies ({\it filled circles\/}), LSB galaxies ({\it
  squares\/}), elliptical galaxies with \Vc\ obtained from \hi\ data
  ({\it diamonds\/}), and elliptical galaxies with \Vc\ obtained from
  dynamical models ({\it crosses\/}) are shown. Data with the same 
  $R_{\it last}/R_{25}$ have been shifted to allow comparison.}
\label{fig:rlast}  
\end{figure}

Our reduced $\chi^2$ is significantly higher than those found
by Ferrarese (2002, $\chi^2_\nu=0.5$ for a sample of 13 spiral
galaxies with \sigmac$>70$ \kms\ and 20 elliptical galaxies) and Baes
et al. (2003, $\chi^2_\nu=0.3$ for a sample of 24 spiral galaxies with
\sigmac$>80$ \kms ).
However, this comparison is affected by the different uncertainties
which characterize the \Vc\ and \sigmac\ measurements of the three
datasets.
In order to allow such a comparison we performed the analysis of the
scatter of the data points. We defined the scatter as
\begin{equation}
s = \sqrt{\frac{\sum_{i=1}^N d_i^2\,w_i}{\sum_{i=1}^N w_i}   } 
\label{eq:scatter}
\end{equation}
with
\begin{equation}
d_i = \frac{a\,\sigma_{c,i} - V_{c,i} + b}{\sqrt{a^2+1}}
\label{eq:distance}
\end{equation}
and 
\begin{equation}
w_i=\frac{1}{\Delta\sigma_{c,i} \Delta V_{c,i}}
\label{eq:weight}
\end{equation}
where $d_i$ and $w_i$ are the distance between the $i-$th data point
and the straight line of coefficients $a=1.32$ and $b=46$ given in
Eq. \ref{eq:HSBfit_linear} and its weight, respectively.
If we consider only the HSB galaxies, the resulting scatter is
$s=11$, $9$ and $23$ \kms\ for \citet{Ferr2002},
\citet{Baes2003} and our sample, respectively.
  The difference in the scatter of the datasets (e.g. [$s$(this
work)$/s$(Ferrarese)]$^2=4.4$) is therefore significantly smaller than
the difference of the corresponding $\chi^2_\nu$
(e.g. $\chi^2_\nu$(this work)$/\chi^2_\nu$(Ferrarese)$ = 5.4$).
This means that the higher value of our $\chi^2_\nu$ is mostly due to
the smaller error bars than to the larger intrinsic scatter of our
HSB+E data points.  
It should be noticed that \citet{Ferr2002} and \citet{Baes2003}
considered only galaxies with a flat rotation curve extending at a
distance $R_{\it last}$ larger than the optical radius $R_{25}$. 
We relaxed this selection criterion to build our final sample and
made sure instead that all rotation curves reached the flat outer
parts. The residual plot of Fig. \ref{fig:rlast} shows that the
scatter of the data points corresponding to our sample galaxies with
\Vc\ measured at $R_{\it last} \geq R_{25}$ is comparable to that of
the galaxies with \Vc\ measured $R_{\it last}<R_{25}$. This confirms
that this particular scale is not important once the asymptotic part
of the rotation curve is reached by the observations. 
  However, Fig. \ref{fig:rlast} indicates that the residuals
are particularly large near $R_{\it last}\simeq R_{25}$ and that the
scatter becomes smaller at $R_{\it last}>1.5 R_{25}$. In the latter
case the flat portion of the rotation curve extends on a larger radial
range and therefore \Vc\ is measured with a higher precision. In fact,
for $R_{\it last}\simeq R_{25}$ the scatter increases symmetrically
with either \Vc$>V_{fit}$ and \Vc$<V_{fit}$ values and it indicates
that the less extended velocity curves are not introducing any
systematic effect. Indeed, the slope of the \vsigma\ relation that we
find is consistent with the one proposed by
\citet{Ferr2002} and
\citet{Baes2003} from a sample of more extended velocity curves.

The measured scatter of the complete sample is $s=18$
\kms, which is   larger than typical measurement errors for
\Vc\ and \sigmac\ ($\simeq 10$\kms ). 
For this reason, the measured scatter is dominated by the intrinsic scatter
that we estimate to be $\simeq 15$\kms .  

We investigated the location of the elliptical galaxies with \Vc\
based on \hi\ data and of LSB galaxies in the \vsigma\ plane. These types
of galaxies were not considered by \citet{Ferr2002} and \citet{Baes2003}.

The data points corresponding to the 5 elliptical galaxies with \Vc\
based on \hi\ data follow the same relation as the remaining disc and
elliptical galaxies. 
For these \hi\ rotation curves we relaxed the flatness criterion in
favor of their large radial extension which is about 10 times larger
than that of optical rotation curves. The inclusion of these data
points does not change the fit based on the remaining disc and
elliptical galaxies.
They are mostly located on the upper end of the
\vsigma\ relation derived for disc galaxies, in agreement with the findings of
\citet{Bert1993}. They studied these elliptical galaxies and showed
that their DM content and distribution are similar to those of spiral
galaxies. 

The LSB and HSB galaxies do not follow the same \vsigma\ relation. In
fact, most of the LSB galaxies are characterized by a higher \Vc\ for
a given \sigmac\ (or a lower \sigmac\ for a given \Vc ) with respect
to HSB galaxies (Fig. \ref{fig:sample}). By applying to the LSB data
points the same regression analysis which has been adopted for the HSB
and elliptical galaxies of the final sample, we find
\begin{equation}
V_c  = (1.35\pm 0.19)\,\sigma_c + (81\pm23)  
\label{eq:LSBfit_linear}
\end{equation}
with \Vc\ and \sigmac\ expressed in \kms . The straight line
corresponding to this fit, which is different from the one obtained
for HSB and elliptical galaxies and happens to be parallel to it, 
is plotted in Fig. \ref{fig:sample}.

\begin{figure} 
\epsscale{.80}
\plotone{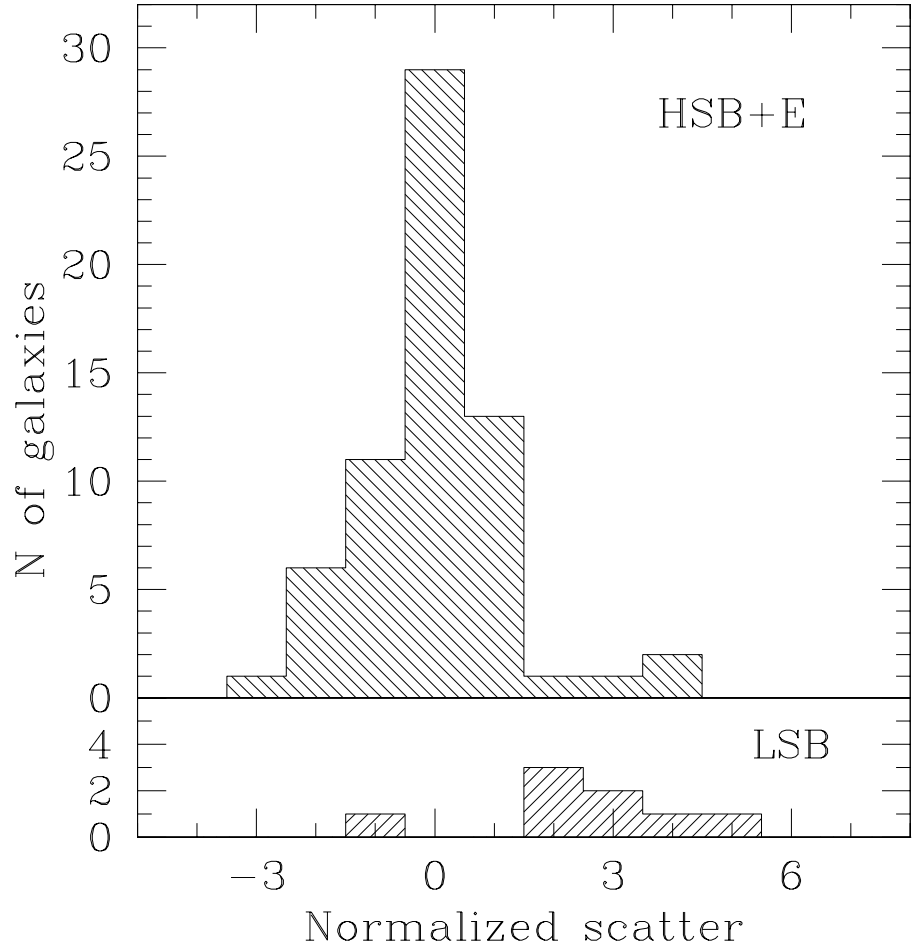}
\caption{The distribution of the normalized scatter 
  of the HSB and elliptical galaxies ({\it upper panel\/}) and LSB
  galaxies ({\it lower panel\/}) with respect to the linear-law fit to
  HSB and elliptical galaxies (Eq. \ref{eq:HSBfit_linear}).  The two
  distributions are different at $>99\%$ confidence level.}
\label{fig:histogram}  
\end{figure}

To address the significance of this result, which is based only on 8
data points, we compared the distribution of the normalized scatter of the
LSB galaxies to that of the HSB and elliptical galaxies.
We defined normalized scatter of the $i-$th data point as
\begin{equation}
\overline{s_i} = d_i / \Delta_i
\label{eq:normalizedscatter}
\end{equation}
where $d_i$ is the distance to the straight line of coefficient
$a=1.32$ and $b=46$ given in Eq. \ref{eq:HSBfit_linear} of the $i-$th
data point, whose associated error $\Delta_i$ is
\begin{equation}
\Delta_i = \sqrt{\Delta V_{c,i}\Delta\sigma_{c,i}}. 
\label{eq:error}
\end{equation}
We assumed $\overline{s_i}>0$ when the data point lies above the
straight line corresponding to the best fit.
In Fig. \ref{fig:histogram} we plot the distributions of the
normalized scatter of the LSB galaxies and of the HSB and elliptical
galaxies.  The two distributions appear to be different, as it is
confirmed at a high confidence level ($>99\%$) by a Kolmogorov-Smirnov
test. The fact that these objects fall in a different region of
the \vsigma\ plane confirms that LSB and HSB galaxies constitute two
different classes of galaxies.

Both demographics of supermassive black holes (SMBH) and study of DM
distribution in galactic nuclei benefit from the \vsigma\ relation.
The recent finding that the mass of SMBHs correlates with different
properties of the host spheroid supports the idea that formation and
accretion of SMBHs are closely linked to the formation and evolution
of their host galaxy. Such a mutual influence substantiates the notion
of coevolution of galaxies and SMBHs \citep[see][]{Ho2004}.

A task to be pursued is to obtain a firm description of all these
relationships spanning a wide range of SMBH masses and address if they
hold for all Hubble types. In fact, the current demography of SMBHs
suffers of important biases, related to the limited sampling over the
different basic properties of their host galaxies.
The finding that the \vsigma\ relation holds for small values of
\sigmac\ points to the idea that SMBHs with masses smaller than
about $10^6$ M$_\odot$ may also exist and follow the
M$_\bullet$-$\sigma$ relation.

Moreover, it has been suggested that the \vsigma\ relation is
equivalent to one between the masses of SMBH and DM halo
\citep{Ferr2000,Baes2003} because \sigmac\ and \Vc\ are related to the
masses of the central SMBH and DM halo, respectively. Yet, this claim
is to be considered with caution, as the demography of SMBHs is still
limited, in particular as far as spiral galaxies are
concerned. 
  Furthermore, the calculation of the virial mass of the DM
halo from the measured \Vc\ depends on the assumptions made for the DM
density profile and the resulting rotation curve \citep[e.g., see the
prescriptions by][]{bull2001,selj2002}. A better estimate of the virial
velocity of the DM halo \Vvir\ can be obtained by constraining the
baryonic-to-dark matter fraction with detailed dynamical modeling of
the sample galaxies. The resulting \Vvir$-$\sigmac\ relation is
expected to have a smaller scatter than the \vsigma\ relation. 
If the M$_\bullet$-$\sigma$ relation is to hold, the
deviation of LSB galaxies {\it with bulge\/} from the \vsigma\ of HSB
and elliptical galaxies suggests that for a given DM halo mass the LSB
galaxies would host a SMBH with a smaller mass compared to HSB
galaxies.
  On the other hand, if the fundamental
correlation of SMBH mass is with the halo circular velocity, then LSBs
should have larger black-hole masses for given bulge dispersion.  
The theoretical and numerical investigations of the processes leading
to the formation of LSB galaxies this should be accounted for

The collapse of baryonic matter can induce a further concentration in
the DM distribution \citep{Rix1997}, and a deepening of the overall
gravitational well in the central regions. If this is the case, the
finding that at a given DM mass (as traced by \Vc ) the central
\sigmac\ of LSB galaxies is smaller than in their HSB counterparts,
would argue against the relevance of baryon collapse in the radial
density profile of DM in LSB galaxies.
Confirming that LSB galaxies follow a different \vsigma\ relation will
highlight yet another aspect of their different formation
history. Indeed, LSB galaxies appear to have a central potential well
less steep than HSB spirals of the same DM halo mass. If the collapse
of baryonic matter causes a compression of the DM halo as well, for LSB
galaxies such process may have been less relevant than for HSB
galaxies. Again LSB galaxies turn out to be the best tracers of the
primordial density profile of DM haloes and therefore in pursuing the
nature of dark matter itself.

\acknowledgments

We are indebted to Matthew Bershady for providing us the BCES code,
which was used to analyze the data.  We wish to thank Maarten Baes and
Laura Ferrarese for stimulating discussion. This research has made use
of the Lyon-Meudon Extragalactic Database (LEDA) and of the NASA/IPAC
Extragalactic Database (NED).

\clearpage



\clearpage


\clearpage

\begin{table*}
\begin{scriptsize}
\begin{center}

\caption{Galaxy Sample}
\begin{tabular}{llcrcrrcc}
\tableline\tableline
\multicolumn{1}{c}{Name} &
\multicolumn{1}{c}{Morp. Type} &
\multicolumn{1}{c}{$i$} &
\multicolumn{1}{c}{$D$} &
\multicolumn{1}{c}{$M_{B_T}^0$} &
\multicolumn{1}{c}{$\sigma_c$} &
\multicolumn{1}{c}{$V_c$} &
\multicolumn{1}{c}{$R_{\it last}$} &
\multicolumn{1}{c}{Ref.}\\
\multicolumn{1}{c}{} &
\multicolumn{1}{c}{} &
\multicolumn{1}{c}{[$^\circ$]} &
\multicolumn{1}{c}{[Mpc]} &
\multicolumn{1}{c}{[mag]} &
\multicolumn{1}{c}{[km s$^{-1}$]} &
\multicolumn{1}{c}{[km s$^{-1}$]} &
\multicolumn{1}{c}{[$R_{25}$]} \\
\multicolumn{1}{c}{(1)} &
\multicolumn{1}{c}{(2)} &
\multicolumn{1}{c}{(3)} &
\multicolumn{1}{c}{(4)} &
\multicolumn{1}{c}{(5)} &
\multicolumn{1}{c}{(6)} &
\multicolumn{1}{c}{(7)} &
\multicolumn{1}{c}{(8)} &
\multicolumn{1}{c}{(9)} \\
\multicolumn{9}{c}{HSB galaxies}\\
\tableline
ESO 323-G25 &(R$'$)SBbc(s):& 55 & 59.8 & $-21.2$ & 139 $\pm$ 14 & 228 $\pm$ 15 & 1.2 & 1\\
ESO 382-G58 &SBbc(r): sp   & 79 &106.2 & $-22.2$ & 165 $\pm$ 22 & 315 $\pm$ 20 & 1.1 & 1\\
ESO 383-G02 & SABc(rs) & 60 & 85.4 & $-21.1$ & 109   $\pm$ 28  & 190   $\pm$ 14  & 1.0 & 1\\
ESO 383-G88 & SABbc(r)?& 67 & 59.5 & $-20.6$ &  70   $\pm$ 14  & 177   $\pm$ 16  & 1.0 & 1\\
ESO 445-G15 & Sbc      & 66 & 60.3 & $-20.2$ & 113   $\pm$ 13  & 190   $\pm$ 21  & 1.1 & 1\\
ESO 446-G01 & SAbc(s): & 53 & 98.3 & $-21.4$ & 123   $\pm$ 12  & 213   $\pm$ 17  & 1.0 & 1\\
ESO 501-G68 & Sbc      & 70 & 45.8 & $-20.1$ & 100   $\pm$ 16  & 173.0 $\pm$ 9.0 & 1.1 & 1\\
IC 342   & SABcd(rs)   & 12 & 2.0  & $-20.5$ &  77   $\pm$ 12  & 185   $\pm$ 10  & 1.4 & 2\\
IC 724   & Sa          & 55 & 78.0 & $-21.6$ & 207.0 $\pm$ 2.8 & 284   $\pm$ 24  & 0.8 & 3\\
NGC 753  & SABbc(rs)   & 39 & 90.0 & $-22.4$ & 121   $\pm$ 17  & 210.0 $\pm$ 7.0 & 0.6 & 2\\ 
NGC 801  & Sc          & 79 & 79.2 & $-21.9$ & 144   $\pm$ 27  & 216.0 $\pm$ 9.0 & 1.6 & 2\\ 
NGC 1160 & Scd:        & 62 & 36.6 & $-21.0$ &  25   $\pm$ 13  & 155.1 $\pm$ 4.0 & 0.8 & 4\\
NGC 1357 & SAab(s)     & 47 & 26.2 & $-20.0$ & 121   $\pm$ 14  & 259.0 $\pm$ 5.0 & 0.8 & 2\\ 
NGC 1620 & SABbc(rs)   & 71 & 46.1 & $-21.1$ &  92.2 $\pm$ 6.9 & 227.5 $\pm$ 4.7 & 1.1 & 5\\ 
NGC 2179 & SA0/a(s)    & 47 & 35.6 & $-19.9$ & 175.6 $\pm$ 6.5 & 241   $\pm$ 11  & 1.0 & 3\\ 
NGC 2590 & SAbc(s):    & 72 & 63.4 & $-21.0$ & 196   $\pm$ 16  & 259   $\pm$ 18  & 1.0 & 5\\ 
NGC 2639 & (R)SAa(r)?  & 53 & 66.5 & $-21.9$ & 195   $\pm$ 13  & 318.0 $\pm$ 4.0 & 0.7 & 2\\ 
NGC 2815 & SBb(r):     & 72 & 40.0 & $-21.6$ & 205.9 $\pm$ 9.8 & 270.7 $\pm$ 5.8 & 0.9 & 5\\
NGC 2844 & SAa(r):     & 61 & 26.4 & $-18.9$ & 113   $\pm$ 12  & 171   $\pm$ 10  & 1.0 & 2\\
NGC 2998 & SABc(rs)    & 63 & 67.4 & $-21.6$ & 113   $\pm$ 30  & 198.0 $\pm$ 5.0 & 1.7 & 2\\
NGC 3054 & SABb(r)     & 52 & 28.5 & $-20.3$ & 138   $\pm$ 13  & 244.5 $\pm$ 3.3 & 0.9 & 5\\
NGC 3038 & SAb(rs)     & 58 & 41.4 & $-21.5$ & 160   $\pm$ 16  & 256   $\pm$ 22  & 1.3 & 1\\
NGC 3145 & SBbc(rs)    & 60 & 61.0 & $-22.1$ & 166   $\pm$ 12  & 261.0 $\pm$ 3.0 & 0.8 & 2\\
NGC 3198 & SBc(rs)     & 68 &  9.4 & $-19.7$ &  69   $\pm$ 13  & 150.0 $\pm$ 3.0 & 5.2 & 2\\
NGC 3200 & SABc(rs):   & 73 & 43.4 & $-21.5$ & 191.9 $\pm$ 3.9 & 272   $\pm$ 12  & 0.9 & 5\\
NGC 3223 & SAb(s)      & 53 & 46.5 & $-22.4$ & 163   $\pm$ 17  & 261   $\pm$ 11  & 0.8 & 2\\
NGC 3333 & SABbc pec sp& 82 & 59.4 & $-21.3$ & 111   $\pm$ 23  & 208   $\pm$ 12  & 1.0 & 1\\
NGC 3885 & SA0/a(s)    & 67 & 22.3 & $-19.4$ & 124.5 $\pm$ 7.5 & 156   $\pm$ 27  & 0.6 & 6\\
NGC 4378 & (R)SAa(s)   & 21 & 43.1 & $-21.0$ & 198   $\pm$ 18  & 308.0 $\pm$ 1.0 & 0.5 & 2\\
NGC 4419 & SBa(s) sp   & 71 & 17.0 & $-19.6$ &  99   $\pm$ 16  & 172.7 $\pm$ 4.5 & 0.4 & 4\\
NGC 4845 & SAab(s) sp  & 76 & 13.1 & $-19.2$ &  80.6 $\pm$ 2.1 & 136   $\pm$ 17  & 0.5 & 3\\
NGC 5055 & SAbc(rs)    & 55 &  8.0 & $-20.5$ & 103.0 $\pm$ 6.0 & 180.0 $\pm$ 5.0 & 2.4 & 2\\
NGC 5064 & (R$'$)SAab: & 63 & 36.0 & $-21.1$ & 188.3 $\pm$ 4.6 & 272.9 $\pm$ 2.8 & 0.9 & 4\\
NGC 6503 & SAcd(s)     & 71 &  5.9 & $-18.7$ &  48   $\pm$ 10  & 116.0 $\pm$ 2.0 & 1.9 & 2\\
NGC 6925 & SAbc(s)     & 75 & 37.7 & $-21.9$ & 190.0 $\pm$ 4.5 & 261.5 $\pm$ 9.1 & 1.0 & 5\\
NGC 7083 & SAbc(s)     & 53 & 39.7 & $-21.5$ & 100.8 $\pm$ 4.4 & 235   $\pm$ 14  & 0.9 & 5\\
NGC 7217 & (R)SAab(r)  & 34 & 21.9 & $-21.2$ & 171   $\pm$ 17  & 241.0 $\pm$ 4.0 & 0.7 & 2\\
NGC 7331 & SAb(s)      & 70 & 14.9 & $-21.5$ & 139   $\pm$ 14  & 239.0 $\pm$ 5.0 & 1.6 & 2\\
NGC 7531 & SABbc(r)    & 67 & 20.9 & $-20.2$ & 108.7 $\pm$ 5.6 & 168.6 $\pm$ 8.4 & 0.7 & 5\\
NGC 7606 & SAb(s)      & 67 & 42.1 & $-22.2$ & 124   $\pm$ 21  & 240.0 $\pm$ 4.0 & 0.9 & 2\\
\tableline
\end{tabular}
\tablecomments{Parameters of the final sample of galaxies. The columns show the
following:
(2) morphological classification from RC3 for HSB and elliptical
galaxies and from ESO-LV for LSB galaxies, except for ESO 534-G20 (NED); 
(3) disc inclination derived for spirals as $\cos^2i=(q^2-q_0 ^2)/(1-q_0^2)$. The
observed axial ratio $q = a/b$ is taken from RC3 and ESO-LV for HSB
and LSB galaxies, respectively, except for ESO 446-G17
\citep{Palu2000}, ESO 206-G14 \citep{Mcga2001}, IC 724
\citep{Rubi1985}, and galaxies from \citet{Baes2003} for which we
adopted their inclination. The intrinsic flattening $q_0 = 0.11$ is
assumed following \citet{guth1992}. For ellipticals with \hi\ data the
inclination is taken from papers listed in column 9;
(4) distance either from papers listed in (10) or derived as $V_0/H_0$
with $H_0=75$ km s$^{-1}$ Mpc$^{-1}$ and $V_0$ the systemic velocity
corrected for the motion of the Sun with respect to the Local Group as
in \citet{RSA}; 
(5) absolute total blue magnitude from $B_T$ corrected for inclination
and extinction from RC3 for HSB and elliptical galaxies and from
ESO-LV for LSB galaxies;
(6) central velocity dispersion of the spheroidal component; 
(7) galaxy circular velocity;
(8) farthest observed radius of the ionized-gas velocity curve in
units of $R_{25}$. $R_{25}$ is from RC3 for HSB and elliptical galaxies and
from ESO-LV for LSB galaxies;
(9) list of references.}
\label{tab:sample}
\end{center}
\end{scriptsize}
\end{table*}

\addtocounter{table}{-1}
\begin{table*}
\begin{scriptsize}
\begin{center}

\caption[Sample]{(continued)} 

\begin{tabular}{llcrcrrcc}
\tableline\tableline
\multicolumn{1}{c}{Name} &
\multicolumn{1}{c}{Morp. Type} &
\multicolumn{1}{c}{$i$} &
\multicolumn{1}{c}{$D$} &
\multicolumn{1}{c}{$M_{B_T}^0$} &
\multicolumn{1}{c}{$\sigma_c$} &
\multicolumn{1}{c}{$V_c$} &
\multicolumn{1}{c}{$R_{\it last}$} &
\multicolumn{1}{c}{Ref.}\\
\multicolumn{1}{c}{} &
\multicolumn{1}{c}{} &
\multicolumn{1}{c}{[$^\circ$]} &
\multicolumn{1}{c}{[Mpc]} &
\multicolumn{1}{c}{[mag]} &
\multicolumn{1}{c}{[km s$^{-1}$]} &
\multicolumn{1}{c}{[km s$^{-1}$]} &
\multicolumn{1}{c}{[$R_{25}$]} \\
\multicolumn{1}{c}{(1)} &
\multicolumn{1}{c}{(2)} &
\multicolumn{1}{c}{(3)} &
\multicolumn{1}{c}{(4)} &
\multicolumn{1}{c}{(5)} &
\multicolumn{1}{c}{(6)} &
\multicolumn{1}{c}{(7)} &
\multicolumn{1}{c}{(8)} &
\multicolumn{1}{c}{(9)} \\
\tableline
\multicolumn{9}{c}{LSB galaxies}\\  
ESO~186-G55 & Sab(r)?    & 63 &  60.1 & $-19.1$ &  91.7 $\pm$ 2.0 & 235   $\pm$ 11  & 1.0 & 7\\
ESO~189-G07 & SABbc(rs)  & 49 &  37.5 & $-20.2$ &  91.3 $\pm$ 2.0 & 185.9 $\pm$ 6.9 & 1.0 & 7\\
ESO~206-G14 & SABc(s)    & 39 &  60.5 & $-19.0$ &  54.3 $\pm$ 2.0 & 141.0 $\pm$ 4.5 & 1.4 & 7\\
ESO~234-G13 & Sbc        & 69 &  60.9 & $-19.3$ &  64.1 $\pm$ 2.0 & 194   $\pm$ 19  & 1.2 & 7\\
ESO~446-G17 & (R)SBb(s)  & 54 &  58.9 & $-20.3$ & 133.6 $\pm$ 2.0 & 196   $\pm$ 33  & 1.2 & 7\\
ESO~450-G20 & SBbc(s):   & 30 &  31.6 & $-19.5$ & 112.4 $\pm$ 2.4 & 245   $\pm$ 35  & 1.0 & 7\\
ESO~514-G10 & SABc(s):   & 36 &  40.4 & $-20.2$ &  60.2 $\pm$ 4.0 & 181   $\pm$ 15  & 0.8 & 7\\
ESO~534-G20 & Sa:        & 46 & 226.7 & $-20.7$ & 153.9 $\pm$ 7.1 & 297   $\pm$ 11  & 1.4 & 7\\
\multicolumn{9}{c}{Ellipticals with $V_c$ from HI data}\\
IC 2006  & (R)SA0$^-$ & 31  & 16.7 & $-18.9$ & 128.0 $\pm$  1.7 & 221 $\pm$ 14 & 2.4 &   8,9\\
NGC 2865 & E3-4       & 65  & 31.2 & $-20.3$ & 169.4 $\pm$  7.0 & 240 $\pm$ 15 & 3.2 & 10,11\\
NGC 2974 & E4         & 55  & 24.0 & $-20.2$ & 254.8 $\pm$  3.8 & 355 $\pm$ 60 & 1.8 &  12,8\\
NGC 4278 & E1-2       & 45  &  7.9 & $-18.5$ & 250.7 $\pm$  7.7 & 326 $\pm$ 40 & 3.3 & 13,14\\
NGC 5266 & SA0$^-$:   & 63  & 37.1 & $-21.4$ & 182.1 $\pm$  9.3 & 270 $\pm$ 40 & 7.6 & 15,16\\
\multicolumn{9}{c}{Ellipticals with $V_c$ from dynamical models}\\
NGC 315  & E$^+$:	& & 69.3 & $-22.3$ & 333   $\pm$ 50  & 569   $\pm$ 59  & 0.8 & 17,18\\ 
NGC 1399 & E1 pec	& & 18.1 & $-20.9$ & 308   $\pm$ 28  & 424   $\pm$ 46  & 0.5 & 17,18\\
NGC 2434 & E0-1 	& & 14.9 & $-19.3$ & 212.6 $\pm$ 1.7 & 331   $\pm$ 42  & 0.8 & 17,18\\ 
NGC 3193 & E2	        & & 17.3 & $-19.5$ & 209   $\pm$ 29  & 303   $\pm$ 25  & 0.4 & 17,18\\ 
NGC 3379 & E1	        & & 10.1 & $-19.8$ & 202   $\pm$ 18  & 259   $\pm$ 23  & 0.6 & 17,18\\ 
NGC 3640 & E3	        & & 15.2 & $-19.7$ & 177.2 $\pm$ 6.8 & 279.2 $\pm$ 8.7 & 0.2 & 17,18\\ 
NGC 4168 & E2	        & & 28.9 & $-20.2$ & 182.4 $\pm$ 5.8 & 287   $\pm$ 21  & 0.4 & 17,18\\ 
NGC 4278 & E1-2         & &  7.9 & $-18.5$ & 250.7 $\pm$ 7.7 & 416   $\pm$ 13  & 0.1 & 13,18\\ 
NGC 4374 & E1	        & & 12.2 & $-20.4$ & 280   $\pm$ 25  & 410   $\pm$ 31  & 0.3 & 17,18\\ 
NGC 4472 & E2	        & & 11.3 & $-20.9$ & 279   $\pm$ 19  & 464   $\pm$ 36  & 0.2 & 17,18\\
NGC 4486 & E$^+$0-1 pec & & 15.5 & $-21.5$ & 351   $\pm$ 19  & 507   $\pm$ 38  & 0.2 & 17,18\\ 
NGC 4494 & E1-2	        & & 17.0 & $-20.6$ & 124   $\pm$ 17  & 261   $\pm$ 25  & 0.2 & 17,18\\ 
NGC 4589 & E2	        & & 28.9 & $-20.6$ & 214   $\pm$ 30  & 333   $\pm$ 22  & 0.3 & 17,18\\ 
NGC 4636 & E0-1	        & & 10.3 & $-19.6$ & 186   $\pm$ 22  & 341   $\pm$ 13  & 0.2 & 17,18\\
NGC 5846 & E0-1	        & & 21.8 & $-20.8$ & 266   $\pm$ 23  & 338.3 $\pm$ 8.8 & 0.8 & 17,18\\
NGC 6703 & SA0$^-$	& & 34.7 & $-20.7$ & 171.6 $\pm$ 1.6 & 222   $\pm$ 29  & 1.0 & 17,18\\
NGC 7145 & E0	        & & 24.5 & $-19.9$ & 133.1 $\pm$ 4.8 & 210   $\pm$ 31  & 0.8 & 17,18\\
NGC 7192 & E$^+$:	& & 36.8 & $-20.6$ & 186   $\pm$ 17  & 270   $\pm$ 18  & 0.4 & 17,18\\
NGC 7507 & E0	        & & 21.6 & $-20.4$ & 236   $\pm$ 15  & 399   $\pm$ 61  & 0.6 & 17,18\\ 
NGC 7626 & E pec:	& & 48.6 & $-21.4$ & 225   $\pm$ 22  & 401   $\pm$ 32  & 1.0 & 17,18\\ 
\tableline
\end{tabular}
\tablerefs{
1: \citet{Baes2003}, 
2: original references can be found in \citet{Ferr2002}, 
3: \citet{Cors1999}, 
4: \citet{vega2001}, 
5: \citet{Pizz2004..17spirals},
6: \citet{Cors2003},
7: \citet{Pizz2004..LSB}, 
8: \citet{Kim1988}, 
9: \citet{Fran1994F},
10: \citet{Jorg1995},
11: \citet{Schi1995},
12: \citet{Beui2002},
13: \citet{Bart2002},
14: \citet{Lees1994},
15: \citet{Caro1993}, 
16: \citet{Morg1997},
17: \citet{Davi1987},
18: \citet{Kron2000}.}
\end{center}
\end{scriptsize}
\end{table*}


\begin{thebibliography}{}

\expandafter\ifx\csname natexlab\endcsname\relax\def\natexlab#1{#1}\fi

\bibitem[{{Akritas} \& {Bershady}(1996)}]{Akri1996}
{Akritas}, M.~G., \& {Bershady}, M.~A. 1996, \apj, 470, 706

\bibitem[{{Baes} {et~al.}(2003){Baes}, {Buyle}, {Hau}, \&
  {Dejonghe}}]{Baes2003}
{Baes}, M., {Buyle}, P., {Hau}, G.~K.~T., \& {Dejonghe}, H. 2003, \mnras, 341,
  L44

\bibitem[{{Barth} {et~al.}(2002){Barth}, {Ho}, \& {Sargent}}]{Bart2002}
{Barth}, A.~J., {Ho}, L.~C., \& {Sargent}, W.~L.~W. 2002, \aj, 124, 2607

\bibitem[{{Beijersbergen} {et~al.}(1999){Beijersbergen}, {de Blok}, \& {van der
  Hulst}}]{Beij1999}
{Beijersbergen}, M., {de Blok}, W.~J.~G., \& {van der Hulst}, J.~M. 1999, \aap,
  351, 903

\bibitem[{{Bertola} {et~al.}(1996){Bertola}, {Cinzano}, {Corsini}, {Pizzella},
  {Persic}, \& {Salucci}}]{Bert1996}
{Bertola}, F., {Cinzano}, P., {Corsini}, E.~M., {Pizzella}, A., {Persic}, M.,
  \& {Salucci}, P. 1996, \apjl, 458, L67

\bibitem[{{Bertola} {et~al.}(1993){Bertola}, {Pizzella}, {Persic}, \&
  {Salucci}}]{Bert1993}
{Bertola}, F., {Pizzella}, A., {Persic}, M., \& {Salucci}, P. 1993, \apjl, 416,
  L45

\bibitem[{{Beuing} {et~al.}(2002){Beuing}, {Bender}, {Mendes de Oliveira},
  {Thomas}, \& {Maraston}}]{Beui2002}
{Beuing}, J., {Bender}, R., {Mendes de Oliveira}, C., {Thomas}, D., \&
  {Maraston}, C. 2002, \aap, 395, 431

\bibitem[{{Bullock} {et~al.}(2001){Bullock}, {Kolatt}, {Sigad}, {Somerville},
  {Kravtsov}, {Klypin}, {Primack}, \& {Dekel}}]{bull2001}
{Bullock}, J.~S., {Kolatt}, T.~S., {Sigad}, Y., {Somerville}, R.~S.,
  {Kravtsov}, A.~V., {Klypin}, A.~A., {Primack}, J.~R., \& {Dekel}, A. 2001,
  \mnras, 321, 559

\bibitem[{{Burstein} \& {Rubin}(1985)}]{Burs1985}
{Burstein}, D., \& {Rubin}, V.~C. 1985, \apj, 297, 423

\bibitem[{{Carollo} {et~al.}(1993){Carollo}, {Danziger}, \& {Buson}}]{Caro1993}
{Carollo}, C.~M., {Danziger}, I.~J., \& {Buson}, L. 1993, \mnras, 265, 553

\bibitem[{{Corsini} {et~al.}(2003){Corsini}, {Pizzella}, {Coccato}, \&
  {Bertola}}]{Cors2003}
{Corsini}, E.~M., {Pizzella}, A., {Coccato}, L., \& {Bertola}, F. 2003, \aap,
  408, 873

\bibitem[{{Corsini} {et~al.}(1999){Corsini}, {Pizzella}, {Sarzi}, {Cinzano},
  {Vega Beltr{\' a}n}, {Funes}, {Bertola}, {Persic}, \& {Salucci}}]{Cors1999}
{Corsini}, E.~M., {Pizzella}, A., {Sarzi}, M., {Cinzano}, P., {Vega Beltr{\'
  a}n}, J.~C., {Funes}, J.~G., {Bertola}, F., {Persic}, M., \& {Salucci}, P.
  1999, \aap, 342, 671

\bibitem[{{Davies} {et~al.}(1987){Davies}, {Burstein}, {Dressler}, {Faber},
  {Lynden-Bell}, {Terlevich}, \& {Wegner}}]{Davi1987}
{Davies}, R.~L., {Burstein}, D., {Dressler}, A., {Faber}, S.~M., {Lynden-Bell},
  D., {Terlevich}, R.~J., \& {Wegner}, G. 1987, \apjs, 64, 581

\bibitem[{{de Blok} \& {McGaugh}(1997)}]{Debl1997}
{de Blok}, W.~J.~G., \& {McGaugh}, S.~S. 1997, \mnras, 290, 533

\bibitem[{{Ferrarese}(2002)}]{Ferr2002}
{Ferrarese}, L. 2002, \apj, 578, 90

\bibitem[{{Ferrarese} \& {Merritt}(2000)}]{Ferr2000}
{Ferrarese}, L., \& {Merritt}, D. 2000, \apjl, 539, L9

\bibitem[{{Franx} {et~al.}(1994){Franx}, {van Gorkom}, \& {de
  Zeeuw}}]{Fran1994F}
{Franx}, M., {van Gorkom}, J.~H., \& {de Zeeuw}, T. 1994, \apj, 436, 642

\bibitem[{{Gerhard} {et~al.}(2001){Gerhard}, {Kronawitter}, {Saglia}, \&
  {Bender}}]{Gerh2001}
{Gerhard}, O., {Kronawitter}, A., {Saglia}, R.~P., \& {Bender}, R. 2001, \aj,
  121, 1936

\bibitem[{{Guthrie}(1992)}]{guth1992}
{Guthrie}, B.~N.~G. 1992, \aaps, 93, 255

\bibitem[{{Ho}(2004)}]{Ho2004}
{Ho}, L.~C., ed. 2004, {Coevolution of Black Holes and Galaxies}

\bibitem[{{Impey} {et~al.}(1996){Impey}, {Sprayberry}, {Irwin}, \&
  {Bothun}}]{Impe1996}
{Impey}, C.~D., {Sprayberry}, D., {Irwin}, M.~J., \& {Bothun}, G.~D. 1996,
  \apjs, 105, 209

\bibitem[{{Jorgensen} {et~al.}(1995){Jorgensen}, {Franx}, \&
  {Kjaergaard}}]{Jorg1995}
{Jorgensen}, I., {Franx}, M., \& {Kjaergaard}, P. 1995, \mnras, 276, 1341

\bibitem[{{Kim} {et~al.}(1988){Kim}, {Jura}, {Guhathakurta}, {Knapp}, \& {van
  Gorkom}}]{Kim1988}
{Kim}, D.-W., {Jura}, M., {Guhathakurta}, P., {Knapp}, G.~R., \& {van Gorkom},
  J.~H. 1988, \apj, 330, 684

\bibitem[{{Kronawitter} {et~al.}(2000){Kronawitter}, {Saglia}, {Gerhard}, \&
  {Bender}}]{Kron2000}
{Kronawitter}, A., {Saglia}, R.~P., {Gerhard}, O., \& {Bender}, R. 2000, \aaps,
  144, 53

\bibitem[{{Lees}(1994)}]{Lees1994}
{Lees}, J.~F. 1994, in Mass-Transfer Induced Activity in Galaxies, proceedings
  of the Conference held at the University of Kentucky, Lexington, April 26-30,
  1993. Edited by Isaac Shlosman. Cambridge: Cambridge University Press, 1994,
  432

\bibitem[{{McGaugh} {et~al.}(2001){McGaugh}, {Rubin}, \& {de Blok}}]{Mcga2001}
{McGaugh}, S.~S., {Rubin}, V.~C., \& {de Blok}, W.~J.~G. 2001, \aj, 122, 2381

\bibitem[{{Morganti} {et~al.}(1997){Morganti}, {Sadler}, {Oosterloo},
  {Pizzella}, \& {Bertola}}]{Morg1997}
{Morganti}, R., {Sadler}, E.~M., {Oosterloo}, T., {Pizzella}, A., \& {Bertola},
  F. 1997, \aj, 113, 937

\bibitem[{{Palunas} \& {Williams}(2000)}]{Palu2000}
{Palunas}, P., \& {Williams}, T.~B. 2000, \aj, 120, 2884

\bibitem[{{Persic} {et~al.}(1996){Persic}, {Salucci}, \& {Stel}}]{pers1996}
{Persic}, M., {Salucci}, P., \& {Stel}, F. 1996, \mnras, 281, 27

\bibitem[{{Pizzella} {et~al.}(2005){Pizzella}, {Corsini}, {Magorrian}, {Sarzi},
  \& {Bertola}}]{Pizz2004..LSB}
{Pizzella}, A., {Corsini}, E., {Magorrian}, J., {Sarzi}, M., \& {Bertola}, F.
  2005, in preparation

\bibitem[{{Pizzella} {et~al.}(2004{\natexlab{a}}){Pizzella}, {Corsini},
  {Bertola}, {Coccato}, {Magorrian}, {Sarzi}, \& {Funes}}]{Pizz2004IAUS}
{Pizzella}, A., {Corsini}, E.~M., {Bertola}, F., {Coccato}, L., {Magorrian},
  J., {Sarzi}, M., \& {Funes}, J.~G. 2004{\natexlab{a}}, in IAU Symposium,
  337--338

\bibitem[{{Pizzella} {et~al.}(2004{\natexlab{b}}){Pizzella}, {Corsini}, {Vega
  Beltr{\' a}n}, \& {Bertola}}]{Pizz2004..17spirals}
{Pizzella}, A., {Corsini}, E.~M., {Vega Beltr{\' a}n}, J.~C., \& {Bertola}, F.
  2004{\natexlab{b}}, \aap, 424, 447

\bibitem[{{Press} {et~al.}(1992){Press}, {Teukolsky}, {Vetterling}, \&
  {Flannery}}]{NUMREC}
{Press}, W.~H., {Teukolsky}, S.~A., {Vetterling}, W.~T., \& {Flannery}, B.~P.
  1992, {Numerical recipes in FORTRAN. The art of scientific computing}
  (Cambridge: University Press, |c1992, 2nd ed.)

\bibitem[{{Rix} {et~al.}(1997){Rix}, {de Zeeuw}, {Cretton}, {van der Marel}, \&
  {Carollo}}]{Rix1997}
{Rix}, H., {de Zeeuw}, P.~T., {Cretton}, N., {van der Marel}, R.~P., \&
  {Carollo}, C.~M. 1997, \apj, 488, 702

\bibitem[{{Rubin} {et~al.}(1985){Rubin}, {Burstein}, {Ford}, \&
  {Thonnard}}]{Rubi1985}
{Rubin}, V.~C., {Burstein}, D., {Ford}, W.~K., \& {Thonnard}, N. 1985, \apj,
  289, 81

\bibitem[{{Sandage} \& {Tammann}(1981)}]{RSA}
{Sandage}, A., \& {Tammann}, G.~A. 1981, in Carnegie Inst. of Washington, Publ.
  635; Vol. 0; Page 0, 0

\bibitem[{{Schiminovich} {et~al.}(1995){Schiminovich}, {van Gorkom}, {van der
  Hulst}, \& {Malin}}]{Schi1995}
{Schiminovich}, D., {van Gorkom}, J.~H., {van der Hulst}, J.~M., \& {Malin},
  D.~F. 1995, \apjl, 444, L77

\bibitem[{{Schombert} {et~al.}(1992){Schombert}, {Bothun}, {Schneider}, \&
  {McGaugh}}]{Scho1992}
{Schombert}, J.~M., {Bothun}, G.~D., {Schneider}, S.~E., \& {McGaugh}, S.~S.
  1992, \aj, 103, 1107

\bibitem[{{Seljak}(2002)}]{selj2002}
{Seljak}, U. 2002, \mnras, 334, 797

\bibitem[{{Vega Beltr{\' a}n} {et~al.}(2001){Vega Beltr{\' a}n}, {Pizzella},
  {Corsini}, {Funes}, {Zeilinger}, {Beckman}, \& {Bertola}}]{vega2001}
{Vega Beltr{\' a}n}, J.~C., {Pizzella}, A., {Corsini}, E.~M., {Funes}, J.~G.,
  {Zeilinger}, W.~W., {Beckman}, J.~E., \& {Bertola}, F. 2001, \aap, 374, 394

\bibitem[{{Whitmore} \& {Kirshner}(1981)}]{Whit1981}
{Whitmore}, B.~C., \& {Kirshner}, R.~P. 1981, \apj, 250, 43

\bibitem[{{Whitmore} {et~al.}(1979){Whitmore}, {Schechter}, \&
  {Kirshner}}]{Whit1979}
{Whitmore}, B.~C., {Schechter}, P.~L., \& {Kirshner}, R.~P. 1979, \apj, 234, 68

\end{thebibliography}
\end{document}